# Optic Disc Segmentation in Fundus Images: From Classical Image Processing and Deformable Models to Modern AI

Buket D. Barkana
Biomedical Engineering Department, The University of Akron, Akron, OH 44325, USA

**Abstract:** Accurate localization and segmentation of the optic disc (OD) are important for retinal image analysis and glaucoma assessment, yet remain challenging due to variations in illumination, pathology, vascular interference, and poorly defined boundaries. This structured methodological review examines the evolution of OD segmentation from classical image-processing and deformable models to contemporary artificial intelligence (AI)-based approaches. A structured literature search and study-selection process was used to identify representative studies spanning major methodological developments. The review first summarizes commonly used fundus-image datasets, then organizes classical methods by principal mechanisms, including intensity and thresholding, histogram and entropy analysis, morphology, geometric and Hough-transform methods, filtering and feature operators, texture- and region-based approaches, and active-contour and level-set models. This paper pays particular attention to the assumptions, strengths, limitations, and complementary roles of these methods in OD localization and boundary delineation. Representative AI approaches are subsequently examined to illustrate the transition from handcrafted features and explicitly defined priors to learned representations, Transformer-based segmentation, boundary- and shape-aware learning, promptable segmentation, and retinal foundation models. Across these methodological generations, several core segmentation principles persist, including region-of-interest localization, multiscale representation, geometric and anatomical constraints, and boundary regularization, although their implementation has shifted from predefined operators to learned modules, losses, prompts, and pretrained representations. The review further identifies boundary ambiguity, anatomical variability, domain shift, and cross-dataset generalization as continuing challenges.

**Keywords:** optic disc segmentation; retinal fundus imaging; deformable models; active contours; deep learning; vision transformers; foundation models

## 1. Introduction

Detecting and segmenting the optic disc (OD) can improve retinal-image analysis by providing an anatomical reference and preventing the bright OD region from being mistaken for lesions or other structures. The OD can share intensity, shape, and size characteristics with exudates and other abnormalities, making automated localization and boundary extraction challenging. One of the most important clinical applications is glaucoma assessment, for which optic nerve head evaluation is central. Glaucoma is a major cause of irreversible visual impairment, and the projected global burden has motivated continued development of automated retinal-image analysis methods [1,2]. OD localization also serves as a reference for vessel analysis, macular localization, and other computer-aided retinal assessment tasks. Figure 1 illustrates retinal fundus anatomy and examples of relatively easy and challenging OD boundaries.

OD detection and segmentation have evolved from classical image-processing and deformable formulations to data-driven segmentation, Transformer architectures, and foundation models. This review examines the methodological progression, with particular emphasis on the classical principles that established the technical basis of OD segmentation and on how these principles persist or are reformulated in contemporary AI-based approaches. Accordingly, this article is presented as a focused methodological review rather than a systematic or scoping review. Representative studies are selected to illustrate major methodological families, technical developments, strengths, limitations, and conceptual transitions rather than to provide an exhaustive systematic evidence synthesis.

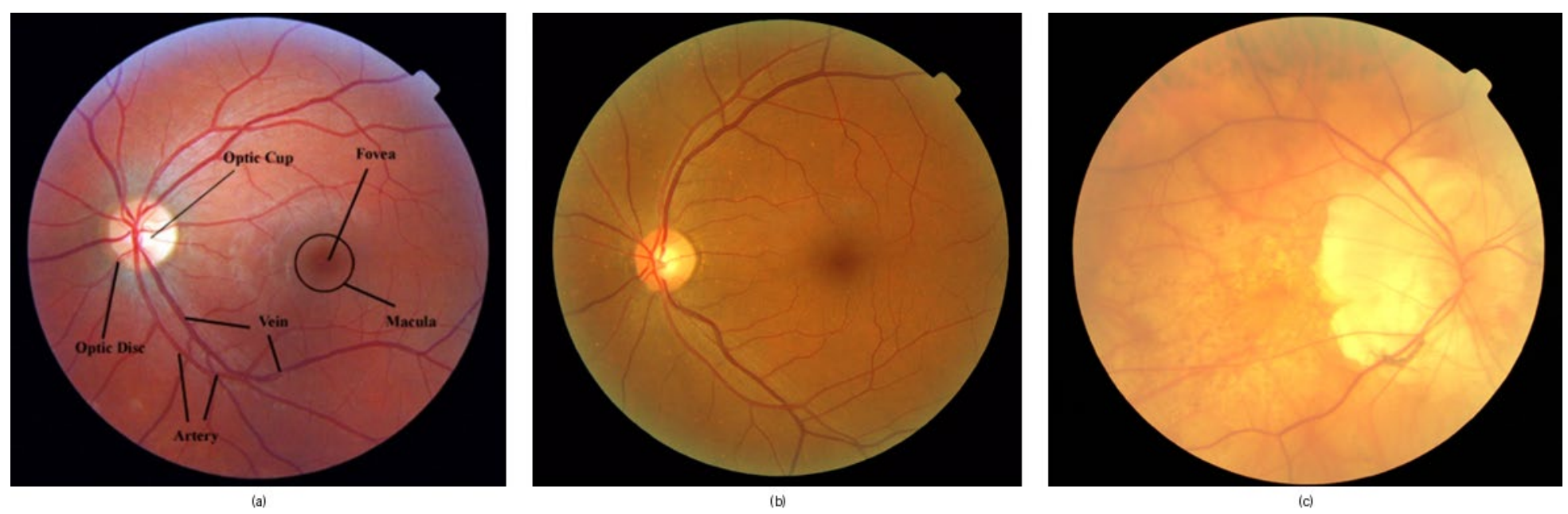

**Figure 1.** Retinal fundus anatomy and examples of OD detection difficulty. (a) Retinal fundus image showing major anatomical features. (b–c) Examples of relatively easy and challenging OD detection cases from the MESSIDOR database [3].

The review first summarizes fundus-image datasets commonly used for OD and optic cup (OC) analysis, then organizes classical methods by primary mechanisms, including intensity and thresholding, histogram and entropy analysis, morphology, geometric and Hough-transform methods, filtering and feature operators, texture- and region-based approaches, and active-contour and level-set models. It examines these methods in terms of their underlying assumptions, processing strategies, strengths, and limitations. The transition to modern AI is then illustrated through selected studies representing convolutional and Transformer-based segmentation, boundary- and shape-aware learning, promptable segmentation, and retinal foundation models. These contemporary approaches are used to demonstrate how classical segmentation principles have evolved into learned frameworks and to examine persistent challenges in boundary delineation, anatomical variability, and cross-dataset generalization.

The remainder of this paper is organized as follows. Section 2 describes the literature search and study-selection process. Section 3 summarizes commonly used fundus-image datasets. Section 4 reviews classical and deformable OD segmentation methods. Section 5 examines the transition to contemporary AI-based approaches. Section 6 discusses methodological continuity, limitations, and generalization, and Section 7 concludes the review.

## 2. Literature Search and Study Selection

A structured literature search was conducted to identify studies relevant to OD image analysis. The screening record compiled for this review covers publications from 2010 through August 2026. Google Scholar served as the literature search platform. Searches combined one of the alternative expressions for the optic disc (“optic disc,” “optic disk,” “optical disc,” or “optical disk”) with at least one image-analysis term (“segmentation,” “localization,” “detection,” “extraction,” “classification,” or “modeling”).

Studies were considered relevant if they (i) reported retinal OD segmentation, localization, detection, extraction, or related computational analysis; (ii) reported the retinal image dataset used in the analysis; and (iii) reported performance metrics. Studies focused only on glaucoma detection without performing OD localization or segmentation were excluded. Studies that mentioned the OD only as an anatomical structure, clinical finding, or disease manifestation, without computational detection or segmentation, were also excluded.

The purpose of this review was to examine methodological developments in optic-disc localization and segmentation rather than to provide an exhaustive comparison of all retrieved studies. Among relevant studies, citation counts available at the time of screening were used as an indicator of scholarly influence when selecting representative studies for detailed methodological discussion. The selected studies were organized by major methodological developments, including classical image-processing approaches, deformable models, and contemporary learning-based methods. For more recent studies, for which citation counts may not yet reflect their impact, methodological novelty and relevance to emerging segmentation approaches were also considered. Foundational methodological references and original dataset descriptions published outside the screening period were included as contextual sources where relevant. Figure 2 presents the documented literature search and study-selection flow.

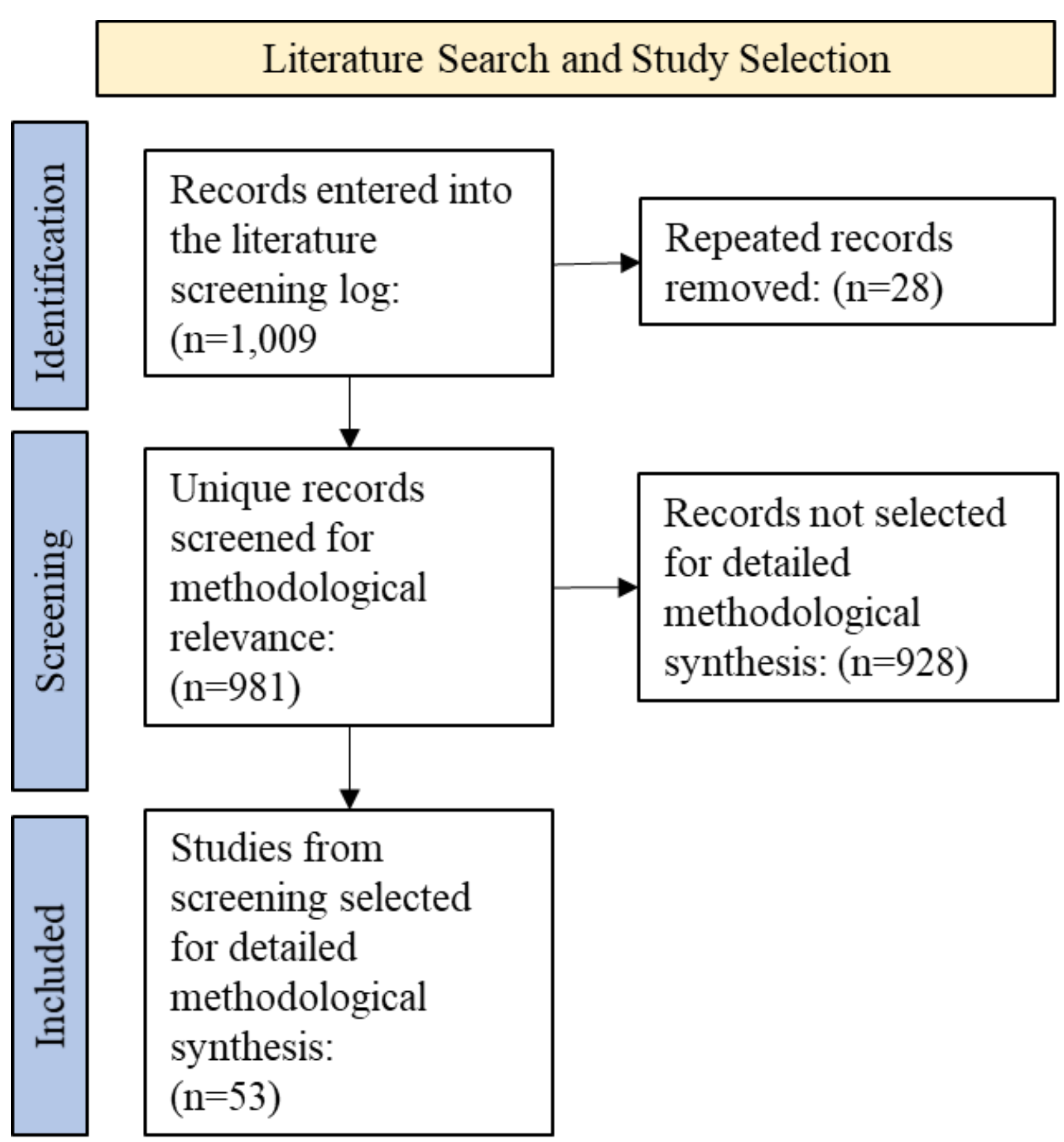

**Figure 2.** Literature search and study-selection flow for the methodological review of optic-disc localization and segmentation studies published between 2010 and August 2026.

## 3. Public Fundus Image Datasets

OD segmentation research has been evaluated on a diverse set of public fundus-image databases, including MESSIDOR [3], DRIVE [4], STARE [5], DIARETDB0 and DIARETDB1 [6,7], DRIONS-DB [8], ONHSD [9], RIGA [10], ARIA [11], HRF [12,24], RIM-ONE DL [13], ORIGA-light [14], IDRiD [15], Drishti-GS1 [16], REFUGE [17], CHASE-DB1 [18], INSPIRE [19], ACRIMA [20], Chákṣu [21], TongjiU-DROD [22], FOVEA [23], and CSDI cataract dataset [25]. These datasets differ in image resolution, camera field of view, pathology distribution, and the availability of OD/OC annotations. Table 1 summarizes the datasets emphasized in this review. Among the fundus-image datasets reviewed, access conditions vary considerably. Several datasets, such as DRIVE, STARE, RIGA, HRF, Drishti-GS1, REFUGE, ACRIMA, Chákṣu, TongjiU-DROD, and FOVEA, are publicly available for research use, although some may require registration or a formal access request. In contrast, some older datasets have limited or uncertain current availability, and others, such as ORIGA-light, may be accessible only through restricted or third-party sources.

**Table 1.** Fundus image datasets, total number of images, and camera settings.

| **Database** | **Reference Year** | **Number of images** | **Camera/acquisition settings** |
|---|---|---|---|
| MESSIDOR [3] | 2014 | 1200 eye fundus color images | A color video 3CCD camera on a Topcon TRC NW6 non-mydriatic retinograph with a 45-degree of view at 1440x960, 2240x1488, or 2304x1536 pixels |
| DRIVE [4] | 2004 | 40 color fundus images | Canon CR5 non-mydriatic 3CCD camera with a 45-degree field of view (FOV) at 768x584 pixels |
| STARE [5] | 2003 | 402 raw retinal fundus images | A 35-degree FOV camera with a resolution of 700×605 pixels |
| DIARETDB0 / DIARETDB1 [6,7] | 2006<br><br>2007 | 130 color fundus images, of which 20 are normal and 110 contain signs of diabetic retinopathy.<br><br>89 color fundus images, of which 5 are normal and 84 contain at least mild non-proliferative signs of diabetic retinopathy | A 50-degree FOV digital fundus camera with unknown camera settings<br><br>50-degree FOV fundus camera with varying imaging settings |
| DRIONS-DB [8] | 2008 | 110 color fundus images | HP-PhotoSmart-S20 high-resolution scanner, RGB format, with a resolution of 600×400 pixels |
| ONHSD [9] | 2004 | 100 fundus images | Canon CR6 45MNf fundus camera, with a 45-degree FOV, and a resolution of 640×480 pixels |
| RIGA [10] | 2018 | Contains three different files: a MESSIDOR dataset file, a Bin Rushed Ophthalmic Center file, and a Magrabi Eye Center file. There is a total of 750 original images and 4500 manually marked images | MESSIDOR: Topcon TRC-NW6 non-mydriatic fundus camera<br><br>Bin Rushed Ophthalmic Center: Canon CR-2 non-mydriatic digital retinal camera<br><br>Magrabi Eye Center: Topcon TRC-50DX mydriatic retinal camera |
| ARIA [11] | 2011 | 143 color fundus images | Camera of unknown specifications; image resolution: 768 × 576 pixels |
| HRF [12]<br><br>HRF-Seg+ [24] | 2013<br><br>2025 | The segmentation dataset contains 15 images of healthy patients, 15 images of patients with diabetic retinopathy and 15 images of glaucomatous patients. Extended OD/OC and peripapillary annotations released as HRF-Seg+ | Canon CR-1 fundus camera with a 45-degree FOV and different acquisition settings |
| RIM-ONE DL [13] | 2020 | 313 retinographies from normal subjects and 172 retinographies from patients with glaucoma | Nidek AFC-210 non-mydriatic fundus camera with a 21.1-megapixel Canon EOS 5D Mark II body, with a FOV of 45 |

| | | | degrees, as well as a non-mydriatic Kowa WX 3D stereo fundus camera |
|---|---|---|---|
| ORIGA-light [14] | 2010 | 650 retinal images of which 168 are glaucomatous images and 482 are non-glaucomatous images | Camera of unknown specifications |
| IDRiD [15] | 2018 | The segmentation file contains a total of 81 images, and the disease grading and localization files both contain a total of 516 retinal images, with 413 images in the training set and 103 images in the test set | Kowa VX-10 alpha digital fundus camera with 50-degree FOV. The images have a resolution of 4288×2848 pixels |
| Drishti-GS1 [16] | 2015 | 101 retinal images with 50 training and 51 testing images | Camera of unknown specifications with a 30-degree FOV and a resolution of 2896×1944 pixels |
| REFUGE Challenge [17] | 2020 | 1200 retinal fundus images, of which 120 are with glaucoma, and 1080 are without glaucoma | Zeiss Visucam 500 fundus camera with a resolution of 2124×2056 pixels (400 images) and a Canon CR-2 device with a resolution of 1634×1634 pixels (800 images) |
| CHASE-DB1 [18] | 2012 | 28 color retina images | NIDEK NM-200D fundus camera with a resolution of 999×960 pixels |
| INSPIRE-stereo / INSPIRE-AVR [19] | 2011 | 30 stereo color images<br><br>40 color images | Fixed-base Nidek 3Dx digital stereo retinal camera with a resolution of 768×1019 pixels<br><br>30-degree FOV Zeiss fundus camera with a resolution of 2392×2048 pixels |
| ACRIMA [20] | 2019 | 705 fundus images, with 396 glaucomatous and 309 normal images | 35-degree FOV Topcon TRC retinal camera and IMAGEnet® Capture System |
| Chákṣu [21] | 2023 | 1345 color fundus images, divided into 1009 training and 336 test images | Three commercially available fundus cameras: Remidio non-mydriatic Fundus-on-Phone (FoP) camera, 2448×3264 pixels (1074 images); Forus 3Nethra Classic non-mydriatic fundus camera, 2048×1536 pixels (126 images); and Bosch handheld fundus camera, 1920×1440 pixels (145 images) |
| TongjiU-DROD [22] | 2024 | 400 annotated fundus-image samples acquired from 147 participants. Fundus images from each eye were acquired using two different cameras; 360 samples were used for training and 40 for validation | Zeiss CLARUS 500 fundus camera, approximately 133° FOV, 3912×3912 pixels; and NES-1000P handheld mydriasis-free portable fundus camera, approximately 40° FOV, 1920×1088 pixels |
| FOVEA [23] | 2025 | 40 preoperative color fundus photographs and 40 corresponding intraoperative retinal fundus frames from 40 patients; the dataset also includes 40 intraoperative video clips | Preoperative images: Topcon 3D OCT-1000 fundus camera. Intraoperative images: Zeiss Lumera 700 ophthalmic microscope with an integrated RESIGHT fundus viewing system |
| CSDI cataract dataset [25] | 2026 | 187 unilateral color fundus images from patients examined at Peking University Third Hospital, collected between April 2023 and February 2024 | Canon CR-2 PLUS AF non-mydriatic digital fundus camera |

## 4. Classical and Deformable Optic Disc Segmentation Methods

The non-learning approaches reviewed in this paper span several distinct methodological families. They include intensity-based [26–30] and threshold-based [31–35] methods; histogram-based [36] and entropy-based [37,38] methods; template-based methods [36,39]; morphological processing [39–46]; geometric and Hough-transform methods [47–53]; matched-filter, feature-operator, and edge-detection methods [53–61]; texture-analysis methods [62]; region-based methods [63,64]; and active-contour/level-set methods [65–70]. Table 2 summarizes the principal methodological families considered in this review, along with their underlying segmentation principles, typical roles, strengths, and limitations.

**Table 2.** Taxonomy of classical and deformable approaches for optic-disc segmentation reviewed in this study.

| Section | Method family | Core principle/information used | Typical role in OD analysis | Main strengths | Common limitations |
|---|---|---|---|---|---|
| 4.2 | Intensity-based methods [26–30] | Exploit the relatively high brightness or characteristic intensity distribution of the OD compared with surrounding retinal tissue | Primarily OD localization; sometimes initial segmentation | Simple, computationally efficient, and easy to interpret | Sensitive to nonuniform illumination, bright lesions, image quality, and pathological changes |
| 4.2 | Threshold-based methods [31–35] | Separate OD and background using global, adaptive, multilevel, or saliency-guided intensity thresholds | OD region extraction and segmentation | Low computational complexity; straightforward implementation | Performance depends on threshold selection and intensity contrast; bright lesions may be confused with the OD |
| 4.3 | Histogram-based methods [36] | Characterize or compare image-intensity/color distributions to identify OD-like regions | Localization and candidate-region identification | Uses regional appearance rather than individual pixels | Sensitive to illumination, color variation, and dataset-dependent intensity distributions |
| 4.3 | Entropy-based methods [37,38] | Use image information content, spectral information, or local entropy to identify regions with distinctive OD characteristics | Primarily localization | Can exploit information beyond simple brightness | Performance depends on image quality, local complexity, and the discriminative value of the selected information measure |
| 4.3 | Template-/appearance-prior methods [36,39] | Compare candidate regions with an expected OD appearance, distribution, or geometric representation | OD localization and approximate boundary estimation | Incorporates prior knowledge of expected OD appearance | Limited when OD appearance differs substantially because of pathology, illumination, or anatomical variability |
| 4.4 | Morphological methods [39–46] | Apply dilation, erosion, opening, closing, reconstruction, and related shape-based operations to enhance | Localization, candidate refinement, segmentation, and post-processing | Effective for structural enhancement and removal of small irrelevant regions; | Strongly dependent on structuring-element size and shape; fixed-scale assumptions may not generalize across image resolutions |

| | | or isolate OD structures | | computationally tractable | |
|---|---|---|---|---|---|
| 4.5 | Geometric, circular, and Hough-transform methods<br><br>[47–53] | Exploit the approximately circular or elliptical geometry of the OD using geometric transformations, circle operators, polar representations, or Hough voting | OD localization and boundary estimation | Provides an explicit and interpretable anatomical shape prior; can tolerate incomplete boundary evidence | Circular/elliptical assumptions may be violated by irregular OD shape, peripapillary atrophy, pathology, or weak boundaries |
| 4.6 | Matched-filter, feature-operator, and edge-detection methods<br><br>[53–61] | Detect directional, convergent, corner, gradient, or edge patterns characteristic of the OD boundary and surrounding structures | Localization and boundary detection | Makes direct use of local structural and boundary information; some operators can integrate evidence over incomplete contours | Vessel crossings, noise, low contrast, and discontinuous OD boundaries can produce competing or weak responses |
| 4.7 | Texture-analysis methods<br><br>[62] | Characterize local spatial texture patterns to distinguish the OD from surrounding retinal tissue | Primarily localization | Exploits information not captured by intensity alone | Depends on feature design, neighborhood scale, and texture stability across datasets |
| 4.7 | Region-based methods<br><br>[63,64] | Form OD regions according to spatial connectivity, similarity, homogeneity, adaptive thresholds, or region-merging criteria | OD segmentation | Can operate without requiring a fully defined boundary and naturally incorporates spatial connectivity | Sensitive to initialization, seed selection, region criteria, intensity heterogeneity, and leakage into neighboring regions |
| 4.8 | Active-contour and level-set methods<br><br>[65–70] | Evolve an explicit or implicit contour by minimizing an energy functional incorporating boundary, region, smoothness, and/or shape information | Precise OD boundary segmentation | Flexible representation of nonrigid boundaries; allows integration of region, edge, and shape constraints | Initialization and parameter sensitivity; local minima; weak boundaries, vessel interference, and heterogeneous regions |

### 4.1 Preprocessing Methods

Preprocessing is an important stage in classical OD localization and segmentation pipelines because variations in illumination, image contrast, vascular structures, and retinal abnormalities can interfere with OD identification. Common preprocessing steps include selecting or transforming color channels, enhancing contrast, reducing noise, and suppressing blood vessels and other structures that may obscure the OD boundary.

Color-channel selection is often used to improve the visibility of the OD and surrounding retinal structures. The green channel is commonly used because it provides good contrast between retinal structures and the background and clearly represents the vascular network [29,71]. However, some studies report that the red channel provides stronger contrast between the OD and the surrounding retina [27,44,69]. Rather than selecting a single channel, Dashtbozorg et al. [57] combined information from the red and green channels to generate a grayscale representation, leveraging the relatively high OD contrast in both channels. Color-space transformation has also been used. For example, Khan et al. [63] enhanced the fundus image using dehazing, converted the cropped OD region to the HSV color space, and selected the V channel for subsequent OD detection.

Contrast enhancement and smoothing are also common preprocessing operations. Gaussian filtering can reduce image noise and suppress fine-scale variations, whereas histogram-based enhancement can increase the distinction between the OD and its surroundings. Contrast-limited adaptive histogram equalization (CLAHE) has been used to improve local contrast [29,71], and adaptive histogram equalization has also been incorporated into active-contour preprocessing [69]. Median filtering is another frequently used operation because it can suppress small lesions and thin vascular structures while producing less blurring than mean filtering. However, its performance depends on the filter size, and large neighborhoods may remove useful edge details [59].

Several methods preprocess the image to reduce vascular interference around or within the OD. Khan et al. [63] detected vessels with a multiscale line detector and removed them using a Laplace-transform-based procedure before applying local adaptive thresholding and region growing. In an active-contour approach, Naqvi et al. [70] used haze removal and Difference-of-Gaussian filtering to enhance the OD region, followed by channel-dependent homogenization and vessel inpainting before contour estimation. Collectively, these preprocessing steps aim to reduce image variability and suppress structures that can interfere with subsequent OD localization or boundary estimation.

### 4.2 Intensity- and Threshold-Based Methods

The OD generally appears as a relatively bright region in color fundus images. Consequently, many classical localization and segmentation methods exploit differences in intensity or contrast between the OD and the surrounding retinal tissue [26–30]. Intensity-based approaches are attractive because of their relatively low computational complexity and because they can substantially reduce the search area before more sophisticated segmentation procedures are applied. However, intensity alone does not uniquely characterize the OD. Exudates, imaging artifacts, and other pathological structures may also appear as bright regions, while uneven illumination or low OD-to-background contrast may weaken the expected intensity difference. Thus, intensity information is commonly combined with vessel, spatial, frequency, or geometric information rather than used as the sole criterion for OD detection.

Several forms of intensity information have been used. These include maximum or mean intensity, local intensity variation, contrast between the OD and its neighborhood, and one-dimensional intensity profiles [26–30]. For example, projection-based processing can reduce a two-dimensional retinal image to one-dimensional representations of brightness and vessel-related features [26], while line-scanning methods analyze intensity variations along selected horizontal and vertical profiles rather than processing every image pixel [29]. Such dimensionality reduction can substantially decrease computational cost. The line-scanning approach in [29] combined spatial intensity profiles with frequency-domain information and performed computations on a limited number of one-dimensional signals. Nevertheless, the authors reported reduced performance in images with substantial intensity variations, artifacts, or retinal pathology, illustrating the fundamental limitation of intensity-based detection. Contrast-based methods similarly exploit the expected difference between the OD and its surroundings; however, low contrast at the disc boundary and inaccurate localization of the OD center can substantially affect segmentation accuracy [30].

Thresholding converts intensity information into discrete regions and is commonly used after intensity or contrast enhancement [31–35]. For a grayscale image $I(x,y)$, a simple binary threshold operation can be expressed as

$$B_T(x,y) = \begin{cases} 0, & if \ \ I(x,y) \leq T \\ 1, & if \ I(x,y) \geq T \end{cases} \tag{1}$$

where $T$ is the selected threshold and $B_T(x,y)$ is the resulting binary image. In OD segmentation, pixels above the selected threshold may represent the bright OD region, although the exact interpretation depends on the selected color channel and preprocessing procedure. A single global threshold is rarely sufficient for fundus images because illumination and contrast vary spatially and because bright pathological regions may have gray levels comparable to those of the OD. Consequently, classical OD methods have employed adaptive thresholding, color multi-thresholding, multilevel thresholding, and threshold selection based on local or structural information [31–35]. Color multi-thresholding, for example, divides the intensity range into more than two classes, allowing different intensity ranges to represent background, disc, cup, or other retinal structures. Noor et al. [31] applied this principle separately in the RGB channels, selecting threshold ranges based on the intensity distributions of the OD and cup.

$$B_T(x,y) = \begin{cases} C_1, & I(x,y) \leq T_1 \\ C_2, & T_1 < I(x,y) \leq T_2 \\ C_3, & I(x,y) > T_2 \end{cases} \tag{2}$$

Otsu's method [72] is one of the most widely used automatic threshold-selection approaches in image segmentation. It determines a threshold from the gray-level probability distribution by maximizing the separation between classes. If an image contains L gray levels and pi denotes the probability of gray level i, the probabilities of the two classes separated by a threshold T are

$$w_0(T) = \sum_{i=0}^{T} p_i \quad \text{and} \quad w_1(T) = \sum_{i=T+1}^{L-1} p_i \tag{3}$$

and their corresponding mean gray levels are given in Eq.(4).

$$\mu_0(T) = \frac{\sum_{i=0}^{T} i p_i}{w_0(T)} \quad \text{and} \quad \mu_1(T) = \frac{\sum_{i=T+1}^{L-1} i p_i}{w_1(T)} \tag{4}$$

The between-class variance is then

$$\sigma_B^2(T) = w_0(T) w_1(T) [\mu_0(T) - \mu_1(T)]^2, \tag{5}$$

and the optimal threshold is selected as

$$T^* = arg \left\{ \max_T \sigma_B^2(T) \right\}. \tag{6}$$

The same principle can be extended to multilevel thresholding by partitioning the gray-level distribution into more than two classes. This is useful when the OD cannot be adequately separated from the background with a single threshold. In practice, thresholding is usually integrated with other operations. Iterative morphological opening and closing can first enhance bright OD candidates before automatic thresholding and subsequent edge or Hough processing [33]. Other approaches combine multilevel

Otsu thresholding with clustering or level-set models [34] or determine the threshold from saliency rather than directly from the global gray-level histogram [35]. The latter approach highlights a broader limitation of conventional thresholding: a fixed or globally optimized threshold may not adequately represent images with uneven illumination, pathology, or artifacts.

Thus, intensity- and thresholding-based methods are computationally efficient and useful for candidate-region extraction and search-space reduction, but their performance is strongly affected by variations in image brightness, pathology, and boundary contrast. For this reason, they are most often used as one component of a multistage OD localization or segmentation framework rather than as stand-alone segmentation methods.

### 4.3 Histogram-, Entropy-, and Template-Based Methods

The histogram of a digital image shows the frequency of occurrence of its gray levels and can be interpreted as an estimate of the image's intensity probability distribution. Histogram-based processing has therefore been used in OD analysis for two related purposes: contrast normalization and enhancement, and characterization of the OD's intensity distribution [36]. Histogram equalization redistributes intensity values to increase an image's effective dynamic range, whereas adaptive and contrast-limited adaptive histogram equalization operate locally to compensate for spatial variations in illumination. These transformations are particularly useful for fundus images because illumination is often nonuniform across the retinal field. Histogram-based normalization can consequently improve subsequent localization, thresholding, vessel extraction, and boundary detection. Dehghani et al. similarly noted that illumination equalization considerably improves contrast for subsequent retinal-image analysis [36].

Histogram information can also be used directly as a statistical representation of the OD. Rather than matching an image template pixel-by-pixel, Dehghani et al. [36] constructed separate template histograms from the red, green, and blue components of representative OD regions. Candidate windows were compared with these channel-specific templates using a histogram-similarity measure of the form

$$C(a,b) = \frac{1}{(1+\sum_i (a_i - b_i)^2)} \qquad (7)$$

where $a_i$ and $b_i$ represent corresponding bins of two histograms. As the histograms become more similar, the squared difference approaches zero and $C$ approaches 1. The correlations obtained from the RGB channels were then combined with different weights. The green channel was assigned greater weight because of its relatively high contrast, whereas the blue channel received less weight because it was more susceptible to noise.

An advantage of histogram-based representations is that they summarize the distribution of pixel intensities without requiring exact spatial correspondence. This can provide tolerance to variations in the OD appearance. However, the same property is also a limitation: histograms do not preserve the spatial arrangement of pixels. Two regions with different structures can therefore have similar intensity distributions. Furthermore, the nonuniform appearance of fundus images, pathology, and illumination differences can modify the expected OD histogram. In [36], histogram matching failed when the expected vascular characteristics were absent near the OD or when pathological regions contained a large number of vessels.

While a histogram describes the complete gray-level distribution, entropy expresses the information contained in that distribution as a single statistical quantity. For an image with gray-level probabilities pi, Shannon entropy can be written as

$$H = -\sum_{i=0}^{L-1} p(i) \log_2 p(i). \quad (8)$$

A uniform gray-level distribution has high entropy, whereas a region dominated by only a few intensity values has lower entropy. In image analysis, smoother regions therefore tend to have lower entropy, while regions with greater intensity variation and structural detail tend to have higher entropy. This property has been exploited for OD localization [37,38]. The rationale is that the OD is not a homogeneous bright region: retinal vessels and neural structures pass through the disc, creating local intensity and texture variations. Muhammed [38] therefore assumed that OD patches should exhibit higher entropy than smoother retinal regions. The image was first divided into relatively large, nonoverlapping regions to identify the approximate maximum-entropy area, then into overlapping windows for finer localization. This use of entropy converts structural complexity into a simple scalar feature and requires relatively little computation. Entropy can also be used for adaptive channel selection rather than directly for OD localization. Martinez-Perez et al. [37] processed the three RGB components and measured Shannon information content within the candidate OD region. The channel with the highest entropy was selected for subsequent OD-center detection using the circular Hough transform. This avoided imposing the same preferred channel on every image: the red channel could provide the highest information content in a well-exposed image, whereas the green channel could become preferable when the red channel was saturated. Their multispectral approach was intended to address variations in channel quality and saturation across fundus images.

Despite its simplicity, entropy does not capture the semantic meaning or explicit spatial organization of image structures. High entropy can also occur in regions with lesions, artifacts, or complex vascular patterns unrelated to the OD. Thus, entropy quantifies statistical information content but does not, by itself, distinguish whether that information arises from the anatomical structure of the OD or from another complex retinal region.

Template-based OD methods incorporate prior information about the expected appearance or geometry of the disc [26,39]. A template may be represented explicitly as an image or geometric shape, or implicitly through statistical or feature-based information. Candidate regions are then evaluated based on their similarity to the expected OD representation. This reduces the number of anatomically implausible candidates and can provide computational advantages when the search space is restricted in advance.

Circular and elliptical representations are particularly attractive because the OD is approximately round, even though its actual boundary may deviate from an ideal geometric shape. For example, Aquino et al. [39] extracted boundary information independently from the red and green channels and approximated the resulting boundaries using the circular Hough transform. Their analysis supported using a circular approximation as a practical compromise among segmentation quality, robustness, and computational efficiency. Related appearance-prior approaches can encode expected OD characteristics without performing conventional template matching. For example, using brightness and vascular-orientation features in [26] substantially reduces the dimensionality of the localization problem and, consequently, the computational demand.

The primary limitation of template-based methods is their dependence on the assumed model. Variations in OD size, shape, pigmentation, peripapillary atrophy, vessel occlusion, and weak or incomplete boundaries can reduce similarity to the expected template. A rigid circular template, for instance, cannot follow irregular boundaries as closely as a deformable model. Nevertheless, template methods remain useful when approximate OD localization or segmentation is sufficient and computational simplicity is important.

**4.4 Morphological Methods**

Mathematical morphology is among the most widely used classical approaches for OD detection and segmentation [39–46]. Rather than relying solely on pixel intensity, morphological processing modifies image structures based on their shape and spatial relationships to a predefined structuring element. For this reason, morphological operators are particularly useful for removing vessels and small artifacts, filling discontinuities in the OD boundary, suppressing structures that differ in size or shape from the OD, and refining candidate regions before subsequent segmentation. In the reviewed literature, morphology was rarely used as an isolated procedure; it was commonly combined with thresholding, edge detection, watershed transformation, Hough-based methods, and active contours [39–46]. Morphology served both as a segmentation strategy and as an intermediate processing stage. The two fundamental morphological operations are dilation and erosion. Let A represent the set of foreground pixels in a binary image and S a structuring element. Dilation can be defined as

$$A \oplus \mathrm{S} = \{a + s | a \in A, s \in S\} \quad (9)$$

which expands the foreground region according to the size and shape of $S$. In OD processing, dilation can connect neighboring boundary segments, close small discontinuities, and enlarge candidate regions. Erosion can be expressed as

$$A \ominus \mathrm{S} = \{z | S_z \subseteq \mathrm{A}\} \quad (10)$$

where $S_z$ denotes the structuring element translated to position $z$. Erosion removes foreground pixels for which the structuring element does not fit entirely within the region, thereby suppressing thin structures, detaching small objects, and shrinking segmented regions. The effect of these operators depends strongly on the shape and size of the structuring element. A disk-shaped element, for example, implicitly favors approximately round structures, whereas a linear element can be used to suppress or detect elongated features such as blood vessels. Thus, morphological processing incorporates prior geometric information without requiring an explicit statistical or deformable shape model. Welfer et al. [40,41] illustrate this principle by using morphological operations and structuring elements of different dimensions to enhance retinal structures, suppress lesions and vessels, and refine the OD region. In their two-stage formulation, structuring-element dimensions were selected based on expected OD and vessel dimensions, demonstrating that these parameters directly influence which anatomical structures are preserved or removed. Dilation and erosion can be combined to form two additional fundamental operators: opening and closing. Opening consists of erosion followed by dilation,

$$A \circ \mathrm{S} = (\mathrm{A} \ominus \mathrm{S}) \oplus \mathrm{S} \quad (11)$$

whereas closing consists of dilation followed by erosion,

$$A \cdot \mathrm{S} = (\mathrm{A} \oplus \mathrm{S}) \ominus \mathrm{S}. \quad (12)$$

Opening removes structures that cannot contain the selected structuring element, making it useful for eliminating small or irregular foreground regions while approximately preserving larger structures.

Closing fills small gaps and holes and can smooth discontinuous boundaries. In fundus images, these properties are particularly relevant because blood vessels cross the OD and can divide otherwise continuous disc regions. Iterative opening and closing have therefore been used to enhance bright OD candidates before thresholding and boundary detection, while other approaches employ closing to reduce vascular interference or opening to regularize the segmented disc [41,43,46].

Another useful morphological representation is the morphological gradient, which emphasizes object boundaries. For a grayscale image I, it is expressed as

$$G_S(I) = \delta_S(I) - \varepsilon_S(I), \quad (13)$$

where $\delta_S(I)$ and $\varepsilon_S(I)$ denote the grayscale dilation and erosion using structuring element S, respectively. Pixels where dilation and erosion differ substantially correspond to strong local intensity transitions. Morales et al. [42] used this principle in conjunction with watershed segmentation and explicitly defined the morphological gradient as the pointwise difference between dilation and erosion.

The watershed transform extends this concept by treating the grayscale or gradient image as a topographic surface. Local minima form catchment basins, while ridges between basins define candidate boundaries. This provides a natural framework for extracting closed regions but introduces a major difficulty: noise and small intensity fluctuations generate numerous local minima and can cause severe over-segmentation. Marker-controlled watershed restricts the transformation to selected minima, but the choice of markers is critical; too few or incorrectly placed markers can lead to under-segmentation. Morales et al. [42] addressed this problem using a stochastic watershed strategy in which repeated realizations with randomly positioned markers were used to estimate the probability of meaningful contours. The approach was further combined with vessel inpainting and geodesic transformations rather than relying on watershed segmentation alone.

Morphological reconstruction offers a more selective alternative to conventional dilation and erosion. Rather than applying an operator freely across the image, reconstruction expands a marker image F while constraining the expansion with a mask M. One iteration of geodesic dilation can be expressed as

$$\delta_M^{(1)}(F) = (F \oplus \mathrm{S}) \cap \mathrm{M}, \quad (14)$$

and the operation is repeated,

$$\delta_M^{(k)}(F) = \delta_M^{(1)}(\delta_M^{(k-1)}(F)), \quad (15)$$

until no further change occurs. The final stable result is the reconstruction of F under M,

$$R_M^{(\delta)}(F) = \delta_M^{(k)}(F), \quad \delta_M^{(k)}(F) = \delta_M^{(k-1)}(F). \quad (16)$$

This constraint allows connected structures associated with the marker while suppressing other structures. Welfer et al. [40] used morphological reconstruction (dilation and erosion) as part of a multistage process to suppress bright pathological regions and extract vascular and OD-related structures. Morphological reconstruction was also later used as the principal image-processing concept for OD and optic-cup segmentation in [45].

A key role of morphology in OD segmentation is reducing vascular interference. Retinal vessels cross both the interior and the boundary of the disc, producing dark structures and strong local gradients that can disrupt an otherwise continuous OD boundary. Morphological closing, erosion, reconstruction, skeletonization, and related operations can suppress these structures or restore boundary continuity before the final segmentation. However, this introduces a tradeoff: aggressive vessel removal can also alter useful OD information. Salazar-Gonzalez et al. [43] noted that morphological approaches used for vessel removal may modify the image and corrupt useful structural information.

Morphology is also useful for repairing boundaries produced by other segmentation operators. Edge detection, for example, frequently yields incomplete OD contours because vessels obscure parts of the boundary or because contrast varies around the disc. In [44], a Sobel detector produced candidate edges after vessel suppression; dilation was then used to connect incomplete boundary segments. Because dilation also thickened the contour and introduced neighboring pixels, skeletonization reduced it to a thinner representation, and hole filling then produced a complete OD region. The final contour was subsequently refined with an active contour. Rather than serving as an independent OD detector, morphology in this case served as a structural correction stage between edge detection and deformable boundary refinement.

Similarly, thresholding and morphological operations are often complementary. Thresholding identifies candidate pixels primarily based on intensity, whereas morphological operations impose spatial and shape constraints on the resulting binary regions. Septiarini et al. [46], for example, combined Otsu thresholding with opening and dilation to remove irregular structures and restore the expected disc size before ellipse fitting. Morphological closing was also used to reduce the influence of vessels during cup segmentation. Aquino et al. [39] likewise combined morphological operations with edge detection and the circular Hough transform, illustrating the substantial overlap between the method categories considered in this review.

Overall, the principal advantage of morphological processing is that it provides explicit control over connectivity, size, and shape without requiring training data. Its main limitation is the same property: the result depends strongly on the selected structuring element, its size, and the sequence of operations. A structuring element that is too small may fail to remove vessels or artifacts, whereas one that is too large may remove portions of the OD or distort its boundary. Several reviewed approaches therefore selected morphological parameters empirically or based on image- or database-dependent estimates of OD and vessel dimensions. These dependencies help explain why morphological methods are generally most effective as components of multistage segmentation pipelines rather than as universal stand-alone rules.

### 4.5 Geometric, Circular, and Hough-Transform Methods

The approximately circular or elliptical shape of the OD provides a strong geometric prior for classical OD localization and segmentation [47–53]. Unlike intensity-based approaches, which identify the OD primarily by its brightness, geometric methods search for image structures that match an expected shape. This can reduce confusion with irregular bright lesions; however, it also makes methods sensitive to deviations from the assumed geometry. The OD is not perfectly circular; its boundary may be interrupted by vessels, and peripapillary atrophy (PPA), lesions, and imaging artifacts can generate competing circular or elliptical structures. These effects explain why geometric operators are generally combined with preprocessing, edge detection, morphological operations, or subsequent boundary-refinement methods rather than used independently. Cheng et al. [47], for example, showed that PPA can cause conventional shape-based segmentation to include tissue outside the true OD boundary. The circular Hough transform (CHT) is one of the most frequently used geometric approaches. A circle with center $(x_c, y_c)$ and radius r satisfies

$$(x - x_c)^2 + (y - y_c)^2 = r^2. \quad (17)$$

This same circular model is explicitly used in OD detection implementations such as Abdullah et al. [49] and Yuningsih et al. [52]. Rather than searching for a circle directly in the image domain, the Hough transform maps candidate edge pixels into the parameter space $(x_c, y_c, r)$ . Conceptually, for a set of detected edge pixels E, the accumulator can be represented as

$$H(x_c, y_c, r) = \sum_{(x_i, y_i) \in \varepsilon} \mathbf{1}(|(x_i - x_c)^2 + (y_i - y_c)^2 - r^2| < \varepsilon, \quad (18)$$

where $\mathbf{1}(\cdot)$ is an indicator function and ε allows for discretization of the parameter space. Candidate circles that receive the most consistent edge votes are considered the most likely OD representations. Thus, the CHT can tolerate an incomplete boundary: the entire circumference need not be visible as long as a sufficient number of boundary pixels support the same circle. Yu et al. [53] noted this advantage when comparing circular Hough fitting with deformable models under weak-contrast conditions. This robustness, however, is highly dependent on the quality of the edge map. Vessel crossings, lesions, PPA, and other image structures create spurious edge points that may produce incorrect maxima in the Hough parameter space. Yu et al. [53] explicitly noted that noise and heterogeneous OD structures can produce incorrect Hough peaks. Similarly, Yuningsih et al. [52] emphasized that poor edge detection directly leads to CHT failure; their approach therefore used grayscale conversion, image complementation, CLAHE, morphological opening, median filtering, and Prewitt edge detection before applying the CHT. This illustrates an important characteristic of Hough-based OD methods: the transform imposes geometric consistency but does not itself determine which image edges actually belong to the OD.

Another important parameter is the range of radii searched by the CHT. If r is unknown, the accumulator becomes three-dimensional, increasing both computational demand and the likelihood of false responses. Restricting r improves efficiency but requires prior knowledge of the approximate OD size. Abdullah et al. [49] addressed this problem by resizing images to a common resolution and experimentally limiting the radius to 29–50 pixels. Zahoor and Fraz [50] instead defined the minimum and maximum CHT radii as approximately 1/30 and 1/10 of the image width, respectively. These strategies reduce the search space but remain tied to image dimensions or empirically selected parameter ranges.

A more physically motivated approach was proposed by Yu et al. [53], who estimated the expected OD size from the camera field of view (FOV) and image resolution rather than averaging manually measured OD diameters from a subset of images. If $A_{FOV}$ is the physical retinal area represented by the FOV and $N_{FOV}$ is the number of pixels within that region, the physical image footprint per pixel is

$$f_{img} = \frac{A_{FOV}}{N_{FOV}} \quad (19)$$

If $D_{OD}$ is the assumed physical OD diameter, then $A_{OD} = \pi \left(\frac{D_{OD}}{2}\right)^2$, and the corresponding OD radius in pixels can be estimated as

$$r_{OD} = \sqrt{\frac{A_{OD}}{\pi f_{img}}} = \sqrt{\frac{\left(\frac{D_{OD}}{2}\right)^2}{f_{img}}} \quad (20)$$

Yu et al. used an average optic nerve head diameter of approximately 1.85 mm and estimated OD radii of 70, 100, and 110 pixels for the three MESSIDOR image resolutions. This formulation is particularly useful because it explicitly links the geometric search scale to image acquisition parameters. Nevertheless, it requires reliable FOV information and still assumes an approximate physical OD diameter, so intersubject anatomical variability remains.

Although circular models are computationally attractive, an elliptical representation can better capture the natural OD shape. A rotated ellipse centered at $(x_c, y_c)$ , with semi-axes lengths $a_e$ and $b_e$ and orientation ϕ, can be represented parametrically as

$$x(t) = x_c + a_e cost\, cos\emptyset - b_e sint\, sin\emptyset,$$

$$y(t) = y_c + a_e cost\, sin\emptyset + b_e sint\, cos\emptyset, \qquad 0 \le t < 2\pi. \quad (21)$$

This formulation follows the constrained elliptical Hough approach of Cheng et al. [47]. Their method used restrictions on ellipse dimensions to reduce the likelihood that PPA would be included within the detected OD. In particular, if the strongest ellipse had an axis ratio suggestive of OD-plus-PPA rather than the disc alone, a second constrained solution was considered. This demonstrates how geometric prior knowledge can be incorporated directly into the Hough parameter space. However, the authors also observed cases in which the constraint was insufficient, requiring additional PPA detection and removal before the elliptical Hough procedure was repeated.

Not all geometric OD methods rely on Hough voting. Lu [48] proposed a circular transformation that evaluates image variations along multiple evenly oriented radial line segments. For a candidate center, transitions along radial directions are examined to identify points of maximum intensity variation; when the candidate is near the actual OD center, these transition points tend to align with the OD boundary. The resulting transformation therefore uses both circularity and boundary contrast to estimate the center and boundary within a common framework. The method reduces the search space and avoids explicit segmentation of the entire vascular tree. Nevertheless, its performance depends on the relationship between radial-line length and OD radius, and failures were reported when the optic-cup boundary produced stronger variation than the OD boundary, when the OD boundary had extremely low contrast, or when the disc deviated substantially from radial symmetry.

A related strategy is to transform the circular OD geometry into a representation that is easier to segment. Under a polar coordinate transformation, a point $(x,y)$ relative to a center $(x_c, y_c)$ can be expressed as

$$\rho = \sqrt{(x - x_c)^2 + (y - y_c)^2}, \qquad \theta = atan2(y - y_c, x - x_c). \quad (22)$$

The inverse transformation is

$$x = x_c + \rho cos\theta, \quad \text{and} \quad y = y_c + \rho sin\theta. \quad (23)$$

A roughly circular boundary around the selected center is therefore transformed from a closed contour in Cartesian coordinates into a simpler radial-distance-versus-angle function. Zahoor and Fraz [50] exploited this property after CHT localization: the circular OD region-of-interest (ROI) was transformed into polar coordinates, effectively straightening the boundary; the transformed region was divided into tiles and processed with morphological reconstruction and adaptive thresholding before being transformed back to Cartesian coordinates and fitted with an ellipse. The approach illustrates how a geometric transformation can simplify the segmentation problem rather than attempting to fit the OD boundary directly in the original image domain.

Geometric information can also be encoded using specialized circular operators. Reza [51] proposed an operator consisting of oriented line and curved segments arranged around each candidate pixel. For orientation $n$, the response can be represented compactly as the mean intensity sampled along the corresponding circular operator $C_n(x, y)$,

$$C_n(x,y) = \frac{1}{|C_n|}\sum_{q\in C_n(x,y)} I(q), \tag{24}$$

where $I(q)$ is the image intensity at point $q$. This is the standard notation for the averaging operation used in the original circle-operator formulation. [51] used eight orientations and analyzed the resulting intensity variation to identify characteristic peak-valley patterns around the OD; candidate regions were then validated using histogram spread. Thus, geometric methods need not explicitly fit a complete circle; they may instead use circular sampling geometry to detect the disc's characteristic spatial pattern.

In summary, geometric and Hough-based approaches provide an efficient way to incorporate explicit prior knowledge of OD shape and scale. The CHT is particularly useful when only portions of the OD boundary are visible, while elliptical models accommodate some departures from circularity. Circular or polar transformations can convert the shape prior into simpler one-dimensional or rectangular representations. Their principal weakness is that the assumed geometry is only approximate. Vessel crossings, PPA, lesions, low boundary contrast, irregular OD shapes, inaccurate center estimation, and database-dependent scale can all violate the model assumptions. Consequently, the most successful geometric methods in this group use geometry primarily to localize the OD or constrain the search space, while additional intensity, morphological, thresholding, region-growing, or deformable-model operations refine the final boundary [47–53]. Abdullah et al. [49], for instance, used CHT to estimate the OD center but used GrowCut to obtain the final boundary, while Yu et al. [53] used estimated OD geometry to initialize a hybrid level-set segmentation rather than treating a fitted circle as the definitive anatomical boundary.

### 4.6 Matched Filters, Feature Operators, and Edge-Detection Methods

Filtering and feature-detection methods extract local image characteristics that distinguish the OD from surrounding retinal structures. In OD analysis, filters have been designed to respond to brightness patterns, vessel orientation, boundary transitions, gradient convergence, corners, and edge structure [53–61]. These approaches overlap considerably: a matched filter may act as a boundary detector, a line operator may encode both intensity and orientation, and corner or convergence filters exploit spatial changes in the image rather than absolute intensity alone. Their main advantage is that they can incorporate local structural information without requiring a complete parametric model of the OD. However, their performance depends strongly on the selected filter scale and orientation, as well as on interference from vessels, lesions, noise, and uneven illumination. For an image I(x,y) and a filter kernel h(u,v), a general linear filter response can be expressed as

$$R(x,y) = \sum_u \sum_v I(x-u, y-v) h(u,v). \quad (25)$$

A matched filter differs from a generic smoothing filter in that the kernel h is designed to match a particular image structure. The response is therefore large when the local image pattern matches the expected structure. When orientation is important, a bank of rotated kernels $h_\theta$ can be used,

$$R_\theta(x,y) = T * h_\theta, \qquad R_\max(x,y) = \max_{\theta \in \Theta} R_\theta(x,y), \quad (26)$$

where $*$ denotes convolution and $\Theta$ is the set of examined orientations. This formulation is useful in retinal images because vessels and OD boundary segments occur at multiple orientations.

Directional matched filtering has been used for OD localization by exploiting the convergence and crossing of the major retinal vessels at the disc. Yu et al. [53] used a Gaussian-shaped matched filter aligned with the dominant vessels crossing the OD region. Their kernel approximated vessel cross-sectional intensity profiles, and OD candidates were evaluated based on the matched-filter response and local contrast. This approach highlights an important distinction: the matched filter need not model the OD itself; it can instead model a structural feature associated with the OD, such as the orientation and profile of its vessels.

A more direct, boundary-oriented formulation was proposed by Dharmawan et al. [61], who tailored a modified Dolph-Chebyshev type-I (MDCF-I) matched filter to the OD boundary. After vessel removal, the ROI was convolved with 12 oriented kernels at 15∘ intervals, and the maximum response across orientations was used to construct OD boundary candidates. The original MDCF-I formulation includes several parameters that control the Dolph-Chebyshev response, window dimensions, orientation, and OD scale. For this review, Eq. (26) conveys the common filtering principle more clearly than reproducing the complete paper-specific kernel. The key methodological point is that multiple oriented responses are combined to recover boundary portions that are not adequately represented by a single filter orientation.

Line operators offer a related yet conceptually distinct approach. Instead of correlating an image with a predefined two-dimensional template, they sample intensity along oriented line segments and characterize how local intensity varies with orientation. Lu and Lim [54] exploited the approximately circular brightness pattern of the OD: intensity variation tends to be greatest along a direction crossing the bright OD region and lower along an approximately orthogonal direction. If $D_i(x,y)$ denotes the image variation measured along the $i$-th oriented line segment through pixel $(x,y)$, the local orientation characteristic can be represented as

$$O(\mathrm{x},\mathrm{y}) = \arg\max_{\mathrm{i}} D_i(x,y). \quad (27)$$

This formulation follows the orientation map used in [54]. The spatial pattern of these dominant orientations around a candidate point is then analyzed to locate the OD. Thus, the method does not simply search for high intensity; it searches for the orientation pattern produced by the disc's radial brightness structure. The line length was scaled to retinal-image dimensions to accommodate resolution differences.

The line-operator approach shows that local structural filters can be more robust than simple maximum-intensity detection. Bright lesions may have intensity values comparable to or higher than those of the OD, but they do not necessarily exhibit the same systematic orientation pattern. Nevertheless, line and matched-filter approaches remain scale-dependent: a filter that is substantially smaller or larger than the expected OD or vessel structure may fail to capture the intended pattern.

Whereas matched filters respond to a predefined local pattern, conventional edge detectors respond to spatial changes in intensity. For the image $I(x, y)$, the gradient can be expressed as

$$\nabla I(x, y) = \begin{bmatrix} I_x \\ I_y \end{bmatrix} = \begin{bmatrix} \frac{\partial I}{\partial x} \\ \frac{\partial I}{\partial y} \end{bmatrix}, \quad |\nabla I| = \sqrt{I_x^2 + I_y^2}. \tag{28}$$

Large gradient magnitudes indicate strong local intensity transitions and potential boundaries. In discrete images, derivatives are approximated using operators such as Prewitt, Sobel, or related finite-difference kernels. Edge detection is attractive for OD segmentation because the disc often shows a change in intensity relative to the adjacent retina. However, the OD boundary is not the only strong edge in the region. Blood vessels cross the disc margin, pathological structures can produce strong gradients, and portions of the true OD boundary may be weak or blurred.

This explains why edge detection is typically embedded within a larger processing pipeline. For example, Mithun et al. [56] first localized the OD from the blue channel using thresholding and morphological processing, then applied edge detection and morphology to refine the boundary region. Similarly, the Prewitt operator was used before the CHT in [52], where the quality of the edge map directly influenced subsequent geometric detection. The fundamental limitation is therefore not merely whether an edge can be detected, but whether the detected edge can be distinguished as the anatomical OD boundary rather than a vessel, lesion, or artifact.

Corners provide another representation of local structural change. Unlike an edge, which generally has a dominant intensity-change direction, a corner exhibits substantial variation in multiple directions. This property can be described using the local second-moment or structure matrix.

$$M(x, y) = \begin{bmatrix} w * I_x^2 & w * (I_x I_y) \\ w * (I_x I_y) & w * I_y^2 \end{bmatrix}, \tag{29}$$

where $w$ is a local weighting window and $*$ denotes convolution. The Harris corner response is commonly defined as

$$R_H(x, y) = \det(M) - k[tr(M)]^2 \tag{30}$$

where k controls the relative influence of the determinant and the trace. Dehghani et al. [55] explicitly used this formulation for OD localization.

The eigenvalues of *M* provide an intuitive interpretation. If both are small, the neighborhood is approximately uniform; if one is large and the other small, an edge is present; if both are large, the neighborhood contains a corner or junction. Dehghani et al. exploited the anatomical observation that retinal vessels originate and branch near the OD, producing a greater concentration of corners and bifurcations in its vicinity. A moving window approximately the size of the OD was therefore used to identify regions with high corner density.

Corner density, rather than individual corner positions, is important here. Vessel crossings and bifurcations create many corner-like structures, and their collective spatial concentration becomes the OD feature. Gui et al. [60] subsequently combined a simplified FAST-type screening stage with an improved Harris calculation to reduce the computational burden of evaluating the corner response at every image pixel. They then localized the OD using the region with the greatest corner density. As with other vessel-related methods, however, pathology and image quality may alter the expected distribution of corner features.

Convergence filters extend gradient analysis by measuring not only the magnitude of local gradients but also whether the gradient vectors converge toward a candidate point. This is particularly relevant for approximately convex structures such as the OD. For a candidate point (*x,y*), a convergence index can be expressed as

$$CI(x,y) = \frac{1}{M}\sum_{q \in G(x,y)} cos\emptyset_q \quad (31)$$

where *G(x,y)* is the filter support region, M is the number of sampled locations, and $\emptyset_q$ is the angle between the local gradient at q and the direction from q toward the candidate point. This formulation represents the fundamental principle of the convergence-index filters used in [57,58]. A high positive response indicates that many local gradients are consistently oriented relative to the candidate center.

The sliding-band filter (SBF) [57] extends this idea by allowing a fixed-width band to move radially along different directions and selecting the position that maximizes the convergence response. This flexibility accommodates departures from a perfectly circular OD and can reduce interference from vessels near the disc center. Dashtbozorg et al. used a low-resolution SBF for OD-center estimation and a higher-resolution SBF for boundary extraction, with parameters adapted to image size and camera FOV. A later super-elliptical convergence-index filter [58] further relaxed the circular-shape assumption by using a support geometry suitable for semi-elliptical convex structures.

Alternative edge detectors have also been explored for OD localization. Alshayeji et al. [59] adapted a gravity-law-based edge detector, in which neighboring pixels are treated as interacting objects and their combined intensity- and distance-dependent influence produces an edge response. The method was preceded by normalization and anisotropic diffusion and followed by masking, thresholding, and candidate selection because lesions, vessels, and illumination variations can also produce strong edge responses.

This approach illustrates the general limitation of edge-based OD detection: detecting strong boundaries does not necessarily identify the anatomical OD boundary. Therefore, additional spatial or structural constraints are typically required.

**4.7 Texture-Analysis and Region-Based Methods**

Texture analysis characterizes the spatial organization of pixel intensities rather than considering intensity values independently. This can be useful for OD detection because the disc contains a characteristic mixture of bright tissue and converging vascular structures, whereas other bright retinal regions may exhibit different local texture patterns. Texture descriptors can therefore provide additional discrimination when intensity-based methods cannot reliably separate the OD from exudates or other bright lesions.

One commonly used texture descriptor is the local binary pattern (LBP). For a center pixel with intensity $I_c$ and P neighboring pixels $I_p$, the LBP value can be given by

$$LBP(x,y) = \sum_{p=0}^{P-1} s\left(I_p - I_c\right)2^p, \qquad s(z) = \begin{cases} 1, & z \geq 0 \\ 0, & z < 0. \end{cases} \tag{32}$$

Thus, the local neighborhood is represented by a binary pattern describing whether surrounding pixels are brighter or darker than the center pixel. LBP provides a compact representation of local spatial structure and is relatively insensitive to uniform illumination changes. Akyol et al. [62] used LBP together with structural similarity information to characterize candidate OD regions. Bright regions were first identified using image processing and Otsu thresholding, and SURF key points were used to reduce the number of regions considered for subsequent texture analysis. LBP features extracted from OD and non-OD samples were further represented using a visual dictionary and classified with a random forest.

The advantage of this approach is that candidate regions are evaluated based on local structural characteristics rather than brightness alone. However, pathological structures can also exhibit substantial texture and generate numerous keypoints. Akyol et al. reported that images containing hard exudates required considerably more processing time because keypoints were detected within diseased regions, and each candidate region had to be analyzed. Thus, texture information can improve discrimination among bright regions, but its effectiveness depends on the selected descriptor and on how candidate regions are generated.

Region-based segmentation takes a different approach. Rather than identifying an OD boundary directly, it forms connected regions whose pixels satisfy a predefined similarity or homogeneity criterion. In conventional region growing, segmentation begins with one or more seed pixels and progressively incorporates neighboring pixels that are sufficiently similar to the current region. A simple intensity-based criterion can be expressed as

$$q \in R \;\; if \;\; q \in N(R) \;\; and \;\; |I(q) - \mu_R| \leq T_R, \tag{33}$$

where $R$ is the growing region, $N(R)$ denotes its neighboring pixels, $\mu_R$ is the current mean intensity of the region, and $T_R$ is the permitted intensity difference. Similar criteria can instead be based on color, gradient, texture, or combinations of image properties. The process terminates when no neighboring pixels meet the selected criterion.

The main advantage of region growing is that it combines spatial connectivity with pixel similarity, avoiding the isolated regions that can result from simple thresholding. Its main limitation is its dependence on seed selection and the growth criterion. An inaccurate seed may produce an incorrect region, while a permissive similarity threshold can allow the region to leak across a weak OD boundary. Conversely, a restrictive threshold may terminate growth before the complete OD has been segmented. These problems are particularly relevant in fundus images because vessels, illumination variations, and pathology can create substantial intensity variation within the OD itself.

Khan et al. [63] combined local adaptive thresholding and region growing after OD localization and vessel removal. The localized OD center served as the initial seed, and neighboring pixels were incorporated based on their intensity difference from the growing region; growth terminated when this difference exceeded a predefined threshold. The resulting connected regions were then evaluated using area and eccentricity, and ellipse fitting was used to regularize the final boundary. This illustrates an important feature of region-growing OD methods: similarity-based growth alone may not ensure that the resulting region is anatomically plausible, so geometric constraints are commonly applied afterward. The method could nevertheless partially segment the OD or include surrounding regions when disease affected the outer disc or when substantial contrast variation occurred within the OD.

Statistical region merging (SRM) follows the same general region-based principle but does not grow a single region outward from a predefined seed. Instead, neighboring image regions are progressively merged when their statistical characteristics are sufficiently similar. The process therefore begins with small regions and constructs progressively larger homogeneous regions according to a merging predicate. Nija et al. [64] applied SRM to group neighboring retinal pixels by intensity similarity and homogeneity, then identified OD candidates from the resulting regions. The statistically merged image was converted to HSV space. Candidate bright regions were identified from the value component, and morphological operations were subsequently used to remove small components and regularize the segmented region. Geometric properties such as area, circularity, eccentricity, and major and minor axis lengths were then used to determine whether circle or ellipse fitting was more appropriate for the final OD boundary.

Overall, texture- and region-based methods provide complementary information. Texture analysis characterizes local spatial appearance, whereas region-based methods exploit similarity and connectivity among neighboring pixels. Both can reduce reliance on simple brightness assumptions, but neither completely eliminates the effects of pathology, nonuniform illumination, vascular interference, or weak OD boundaries. Consequently, the reviewed approaches combine these representations with thresholding, morphological operations, geometric constraints, or classification rather than relying on texture or region similarity alone.

**4.8 Active-Contour and Level-Set Methods**

Active-contour models (ACMs), also called snakes, are deformable curves that evolve toward an object boundary by minimizing an energy functional [65–70,73]. Unlike circular or elliptical template methods, an active contour is not constrained to a fixed geometric shape and can adapt to local variations in the OD boundary. This flexibility is particularly useful in fundus images, where the OD may deviate from circularity and its margin may be partially obscured by vessels, pathology, or nonuniform illumination. Active contours can incorporate edge, intensity, texture, and shape information and produce smooth, continuous boundaries. However, their performance depends on image-derived forces, contour initialization, regularization parameters, and the extent to which interfering structures are suppressed. Mary et al. [69] also emphasize the flexibility of ACMs and compare several formulations designed for weak boundaries, heterogeneous regions, noise, and intensity inhomogeneity.

A conventional parametric active contour can be represented by a curve

$$C(s) = [x(s), y(s)], \quad 0 \le s \le 1, \tag{34}$$

whose energy may be expressed in the general form

$$E_{ACM}(C) = \int_0^1 \left[\frac{\alpha}{2}|C'(s)|^2 + \frac{\beta}{2}|C''(s)|^2 + E_{img}(C(s))\right] ds. \tag{35}$$

The first two terms constitute the internal energy of the contour. The first derivative controls contour continuity or tension, whereas the second derivative penalizes excessive curvature and therefore controls smoothness. The external image term $E_{img}$ attracts the contour toward image features such as edges. Minimizing the total energy involves a compromise: the image forces pull the contour toward the OD boundary, while the internal forces prevent the curve from becoming overly irregular. The original snake formulation is therefore particularly suitable when the desired boundary produces a sufficiently strong image response.

This dependence on edge information is also a significant limitation. Blood vessels crossing the OD generate strong gradients that can compete with the true disc margin, while portions of the OD boundary may have very low contrast or merge into the surrounding retina. In such cases, a classical edge-driven snake may stop at an incorrect structure or fail to reach the actual boundary. Gradient vector flow (GVF) active contours partially address this problem by diffusing gradient information away from the edges, thereby increasing the capture range of the external force and improving contour evolution toward concave or poorly initialized boundaries. Mary et al. [69] initialized a GVF-based contour from a circular Hough estimate and allowed the snake to deform iteratively toward the OD boundary as the energy decreased. An alternative is to represent the evolving contour implicitly using a level-set function $\phi(x,y)$. The contour is then defined as the zero level set

$$C = \{(x, y) \in \Omega: \emptyset(\mathrm{x}, \mathrm{y}) = 0\}, \quad (36)$$

where $\Omega$ denotes the image domain. Pixels with positive and negative values of $\phi$ represent the regions on opposite sides of the contour. Rather than moving individual contour points explicitly, the segmentation is obtained by evolving $\phi$. This representation allows the contour to deform naturally while maintaining a closed boundary and provides a convenient framework for incorporating regularization and region-based image information. Joshi et al. [66], for example, represented the OD contour as the zero level set of a function and included distance and contour-length regularization during optimization. Early OD work using implicit geometric active contours similarly combined vessel suppression with level-set evolution initialized near the localized OD region [65].

A major development in active-contour segmentation was the shift from purely edge-driven models to region-based models. Rather than requiring a strong gradient at the boundary, region-based models compare image characteristics inside and outside the evolving contour. A commonly used formulation is the Chan-Vese energy [74],

$$\begin{aligned} E(C, c_{in}, c_{out}) &= \mu L(C) + \vartheta A(C) \\ &\quad + \lambda_{\mathrm{in}} \int_{inside(C)} [I(x, y) - c_{in}]^2 dxdy + \lambda_{\mathrm{in}} \int_{outside(C)} [I(x, y) - c_{out}]^2 dxdy, \end{aligned} \quad (37)$$

where *L(C)* and *A(C)* denote the contour length and enclosed area, respectively; $c_{in}$ and $c_{out}$ represent the mean intensities inside and outside the contour; and the remaining parameters control the relative contributions of the terms. The contour evolves toward a configuration that produces relatively homogeneous regions while maintaining a smooth boundary. Naqvi et al. [70] explicitly use this formulation for OD contour estimation.

The important advantage of $E(C, c_{in}, c_{out})$ is that a distinct edge is not required for the contour to stop. Instead, the boundary can be inferred from differences between the regions on either side. This makes region-based models attractive for OD boundaries that are smooth, partially missing, or submerged in the background. For example, Naqvi et al. [70] selected a gradient-independent variational active contour because the OD rim may be extremely smooth or discontinuous due to vessels and peripapillary atrophy. The simple global region model nevertheless assumes that the interior and exterior can be characterized adequately by global statistics such as $c_{in}$ and $c_{out}$.

Fundus images frequently violate this assumption because of illumination variation, vessel structures, pathology, and heterogeneous OD appearance. Localized and region-scalable models address this problem by estimating image statistics within neighborhoods around individual contour points rather than across the entire image. Joshi et al. [66] incorporated local intensity, color, and texture information into a multidimensional feature space, so that contour evolution was driven by the neighborhood of each point. The purpose was specifically to recover local boundaries that may be missed when the global statistics inside and outside the contour are similar. Malek et al. [67] similarly used a region-scalable fitting formulation with local fitting functions and level-set regularization, targeting boundary leakage and intensity inhomogeneity.

Although active contours are deformable, initialization remains important. The initial curve should generally be sufficiently close to the OD to prevent convergence to an unrelated retinal structure. For this reason, active-contour segmentation is commonly preceded by one of the localization approaches discussed earlier. Joshi et al. [66] used a circular Hough transform after vessel suppression to estimate the OD center and radius, then used the resulting circle to initialize the active contour. Mary et al. [69] similarly used a CHT-derived circle as the initial contour when comparing different ACM formulations. Thus, geometric methods and deformable models are often complementary: the former provide an approximate location and shape, while the latter refine the boundary.

Vessel removal or inpainting is also particularly important because vessels emerging from the OD can fragment the apparent boundary and create gradients stronger than the actual disc margin. Joshi et al. [66] suppressed and inpainted vessel regions before contour initialization. Giachetti et al. [68] separated vessel and OD information and used a vessel-inpainted grayscale image, together with vessel-density information, in a coarse-to-fine segmentation strategy. Their contour model increased in complexity, progressing from an initial circle to an ellipse and finally to a free-form snake used for relatively small local corrections. Naqvi et al. [70] similarly homogenized the OD by suppressing vascular structure before applying the region-based contour. These approaches demonstrate that deformability alone does not solve vessel interference; preprocessing remains important for controlling the image forces that drive contour evolution.

The comparative study by Mary et al. [69] further illustrates that no single active-contour formulation addresses all OD characteristics equally well. Ten ACMs were evaluated, including geodesic, GVF, Chan–Vese, localized region-based, B-spline, distance-regularized, and adaptive diffusion formulations. In their RIM-ONE experiments, the GVF-based Xu-ACM produced statistically lower segmentation errors than the nine alternative ACMs and yielded particularly strong OD segmentation in 94% of the evaluated images. However, PPA and image overexposure remained important failure conditions. The comparison also demonstrates the tradeoffs among ACM formulations: localized models can better accommodate heterogeneous image statistics but are more sensitive to initialization, while methods designed for intensity inhomogeneity, regularization, or weak boundaries do not necessarily provide superior OD segmentation in every dataset.

Initialization can also interact with energy formulation in unexpected ways. Naqvi et al. [70] reported failures for relatively small optic discs when the initialized ellipse completely enclosed the true OD. Because their contour was driven by the balance of inside and outside regional energies rather than explicit edge information, optimization could terminate before the contour contracted to the actual boundary. This illustrates a general limitation of deformable models: greater flexibility does not eliminate dependence on initialization, energy design, and stopping criteria.

Active-contour and level-set approaches offer a more flexible representation of the OD boundary than thresholding, morphology, or fixed geometric models. Edge-based contours exploit boundary gradients but are vulnerable to weak edges and vascular interference; region-based contours can detect boundaries without strong gradients but may be affected by intensity inhomogeneity and inappropriate regional statistics; localized formulations improve adaptation to heterogeneous images at the cost of increased parameter and initialization sensitivity. Level-set representations provide a convenient implicit framework for evolving and regularizing these contours. Across the reviewed OD methods [65–70], the most robust strategies therefore combine deformable boundary estimation with OD localization, vessel suppression or inpainting, multiscale processing, and appropriate region or shape constraints.

## 5. Transition from Classical Foundations to Modern AI

The transition from classical OD segmentation to modern AI primarily involved replacing hand-engineered feature extraction and rule-based decision processes with data-driven feature learning and end-to-end segmentation. However, core segmentation principles, such as multiscale representation, geometric and shape constraints, boundary modeling, and ROI localization, remain embedded in many modern architectures through learned modules, auxiliary losses, coordinate transformations, and coarse-to-fine processing. The studies reviewed in this section were selected as representative methodological milestones illustrating this progression rather than as an exhaustive survey of deep-learning-based OD segmentation.

### 5.1. From Handcrafted Features to Learned Representations

Classical OD segmentation represents retinal structures using predefined descriptors derived from intensity and color [26,27], edge and vascular information [39,43], geometric characteristics [48], and texture [62]. In deep-learning-based segmentation, these descriptors are replaced by hierarchical feature representations learned directly from annotated fundus images. Feature extraction and segmentation can therefore be jointly optimized rather than implemented as separate stages.

Fu et al. [75] implemented this approach using M-Net for joint OD and OC segmentation. The architecture consists of a multiscale input layer, a U-shaped convolutional network, side-output layers, and a multi-label loss. The multiscale input constructs an image pyramid to provide receptive fields at multiple spatial scales, while the convolutional encoder learns hierarchical representations from the input image. Side outputs provide intermediate predictions at multiple depths, and a Dice-based multi-label objective addresses joint OD/OC pixel classification.

Fu et al. [75] also applied a polar transformation centered on the localized OD before segmentation. Transforming the approximately circular OD/OC structures from Cartesian to polar coordinates introduces a spatial constraint and alters their representation, thereby facilitating joint OD/OC segmentation. Thus, although learned representations replace handcrafted appearance features, the pipeline retains classical principles of geometric transformation and multiscale processing.

Yi et al. [76] integrated classical geometric localization with learned segmentation in C2FTFNet, a two-stage coarse-to-fine framework. The coarse stage combines U-Net with the Circular Hough Transform (CHT) to localize the OD and define the region of interest. TransUNet3+ then processes the resulting ROI to perform joint OD/OC segmentation. The Transformer module captures long-range dependencies, while multiscale dense skip connections fuse low-level spatial information with high-level semantic features.

These studies illustrate two mechanisms for integrating classical image-processing principles with learned segmentation. In Fu et al. [75], hierarchical feature learning is combined with an explicit polar coordinate transformation, whereas Yi et al. [76] retain the Circular Hough Transform for OD localization before neural-network-based OD/OC segmentation. In both approaches, learned representations replace manual feature engineering, while geometric localization, multiscale processing, and spatial constraints remain explicit components of the segmentation pipeline.

### 5.2. Transformer-Based Segmentation and Global Context

CNN-based segmentation learns hierarchical image representations through successive convolutional operations; however, it captures long-range spatial relationships indirectly by increasing receptive fields. Transformer architectures use self-attention to model interactions among spatially separated features and capture global context. In C2FTFNet, Yi et al. [76] incorporated a Transformer module into the fine-segmentation stage to model long-range dependencies, while multiscale dense skip connections integrate low-level spatial features with high-level semantic representations. The framework therefore retains coarse localization and multiscale processing while leveraging learned local and global representations for OD/OC segmentation.

Bazi et al. [77] investigated Transformer-based OD/OC segmentation using Vision Transformer (ViT) and hierarchical Swin Transformer backbones. Their framework employs two segmentation heads in a coarse-to-fine configuration. The first head generates an initial OD/OC prediction that defines an OD-centered crop, and the second head processes the cropped region to refine the segmentation. The authors evaluated cross-dataset generalization across eight retinal datasets using a leave-one-out protocol. This two-stage design parallels the localization-and-segmentation structure of several classical approaches discussed in Sections 4.5-4.8, where geometric, region-based, or deformable methods first restrict the OD search region before refining its boundary. In Bazi et al. [77], however, both localization and boundary refinement are performed through learned Transformer representations rather than through explicitly defined geometric or contour-evolution rules.

Wang et al. [22] proposed ODFormer, a Swin Transformer-based framework for optic nerve head segmentation. The architecture incorporates a multiscale context aggregator to capture information across receptive fields and a bidirectional feature recalibration module to improve feature integration. The model was evaluated on fundus datasets acquired with different cameras and under varying imaging conditions to assess cross-dataset generalization. Although self-attention enables modeling of long-range spatial dependencies, the authors noted limitations in preserving local spatial priors, structural information, and fine details during multiscale reconstruction. These limitations parallel the challenges addressed by the classical multiscale, edge-, and structure-based methods described in Sections 4.5 and 4.6. ODFormer therefore replaces manually designed filters and geometric operators with learned multiscale and attention-based representations, while still requiring the integration of global context with local structural information for accurate OD boundary delineation.

These studies show that Transformer-based OD/OC segmentation extends feature modeling beyond predominantly local convolutional operations to long-range spatial interactions. However, accurate segmentation still depends on preserving fine structural and anatomical information. Thus, the main methodological shift is not the removal of classical segmentation principles but their implementation through learned representations and architectural modules rather than through manually defined operators.

### 5.3. Persistence of Boundary, Shape, and Anatomical Priors

Boundary and shape information remains important in contemporary OD segmentation. Classical edge operators, Hough-based methods, active contours, and level sets explicitly incorporate this information via gradient responses, geometric constraints, or contour-evolution terms. In deep-learning-based approaches, segmentation is typically formulated as a pixel-wise classification [75,79]; however, mask prediction alone may not preserve contour continuity, shape regularity, or topological consistency.

He et al. [78] incorporated explicit boundary information into an unsupervised domain-adaptation framework for OD/OC segmentation. Their mask-boundary U-Net (MBU-Net) jointly predicts segmentation masks and boundaries and is combined with self-ensembling and output-level adversarial adaptation to reduce domain differences across retinal datasets. Ablation experiments showed that including the boundary branch improved segmentation performance relative to the corresponding U-Net backbone, demonstrating the contribution of explicit boundary supervision to learned OD/OC segmentation.

Chen et al. [79] provide an even more direct link to classical shape-constrained segmentation. They observed that pixel-level neural segmentation may produce contours that do not preserve the expected geometry or topology of the OD and OC. Their framework therefore combines a learned pixel-level information aggregation network with a contour reconstruction network. Fourier coefficients parameterize the OD and OC contours in polar coordinates, allowing continuity and topological constraints to be incorporated into the reconstructed segmentation. The authors explicitly describe the contour parameters as incorporating prior knowledge because the reconstructed regions remain connected and preserve contour continuity.

In classical deformable models, regularity and shape constraints are imposed explicitly via energy terms, contour parameterizations, or geometric assumptions. In contemporary deep models, similar goals can be achieved via boundary losses, auxiliary contour branches, coordinate transformations, or learned shape-reconstruction modules. Therefore, contemporary learning-based segmentation does not remove explicit prior information; instead, it incorporates it through architectural constraints, auxiliary branches, loss functions, or parameterized contour representations.

### 5.4. Foundation Models and Their Adaptation for OD/OC Segmentation

Foundation models differ from task-specific segmentation networks by learning transferable representations through large-scale pretraining and then adapting them to downstream applications. For OD/OC segmentation, two relevant directions have emerged: general-purpose promptable segmentation models adapted to medical and fundus images, and retinal-specific foundation models for dense anatomical segmentation. Compared with the classical approaches in Section 4, this paradigm shifts feature construction from predefined image descriptors to pretrained representations, while task-specific spatial information is incorporated through prompts, adapters, or segmentation decoders.

Kirillov et al. [80] introduced the Segment Anything Model (SAM), a promptable segmentation foundation model comprising an image encoder, prompt encoder, and mask decoder. The image encoder generates a reusable image embedding, while points, bounding boxes, or masks encode spatial information about the target object. The mask decoder combines the image and prompt embeddings to produce the segmentation mask. SAM was trained on the SA-1B dataset, which contains more than one billion masks from approximately 11 million images, to support zero-shot transfer to previously unseen segmentation tasks. Spatial prompting is relevant to OD segmentation because it provides target-location information analogous to the ROI, seed, or contour initialization used in the geometric, region-based, and deformable methods described in Sections 4.5–4.8; however, SAM predicts the final mask from learned representations rather than explicit region criteria or contour-evolution equations.

Ma et al. [81] adapted SAM for medical image segmentation via MedSAM, using 1,570,263 image-mask pairs across multiple imaging modalities. MedSAM retains prompt-conditioned segmentation while adapting the model to medical imaging characteristics. The authors reported that the original SAM struggled with low-contrast structures or indistinct boundaries, whereas MedSAM improved segmentation under these conditions. Thus, large-scale general-purpose pretraining alone does not eliminate domain-specific segmentation difficulties, particularly when target boundaries are poorly defined.

Wang et al. [82] extended the prompt-based framework to OD/OC segmentation using OSAM-Fundus, a training-free one-shot method that uses a single annotated reference image. Automatic Prompt Generation identifies corresponding regions between the reference and target fundus images through patch-level bidirectional feature matching, supplemented by medical prior information and hard-negative sampling. The generated prompts are provided to SAM, and a subsequent feature re-matching stage refines the initial segmentation and suppresses false-positive regions. The authors further compared point, box, and combined point-and-box prompts and reported that composite prompting generally produced better segmentation. In this approach, target localization is no longer determined by a fixed geometric operator or a manually specified initialization but is inferred from learned feature correspondence between retinal images.

A complementary foundation-model strategy uses retinal-specific pretraining rather than a general-purpose segmentation model. Zhou et al. [83] introduced RETFound, a retinal foundation model pretrained via self-supervised learning on approximately 1.6 million retinal images, including color fundus photographs and optical coherence tomography images. Using a masked autoencoder objective, the model learns retinal representations by reconstructing heavily masked image inputs without task-specific annotations. The original RETFound framework was developed primarily for disease diagnosis and prognosis rather than for OD/OC segmentation, but it established a transferable retinal representation for subsequent downstream adaptation.

Zhao et al. [84] adapted RETFound for joint OD/OC segmentation using FunduSegmenter. RETFound serves as the pretrained encoder, while a pre-adapter, segmentation decoder, post-adapter, attention-based skip connections, and a Vision Transformer block adapter transform and integrate the pretrained features for dense pixel-level prediction. The framework was evaluated through internal, external, and domain-generalization experiments across multiple fundus datasets, assessing the transferability of retinal foundation-model representations to OD/OC segmentation.

These studies mark a shift from task-specific feature learning to reusable pretrained representations. SAM-based methods introduce prompt-conditioned target specification, whereas RETFound-based approaches provide retinal-specific representations that are then adapted for segmentation. In both cases, pretrained representations reduce reliance on manually designed features, but OD/OC segmentation still requires mechanisms for target localization, spatial reconstruction, and accurate boundary definition.

## 6. Discussion

The main methodological shift in OD segmentation is from explicitly defined features and prior constraints in classical methods to data-driven feature learning in AI-based approaches. Classical methods define the relevant image evidence in advance using intensity, morphology, geometry, edges, regional statistics, or contour regularization. Learning-based methods increasingly estimate this evidence from training data, but the segmentation problem still requires decisions about where the OD is located, which image structures define its boundary, and what constitutes an anatomically plausible region.

The deformable-model studies in Section 4 show that accurate OD segmentation depends not only on contour evolution but also on localization, preprocessing, and boundary information. Joshi et al. [66] and Mary et al. [69], for example, initialized active contours using Circular Hough Transform estimates, while Joshi et al. [66], Giachetti et al. [68], and Naqvi et al. [70] used vessel suppression or inpainting to prevent vascular gradients from dominating the OD boundary. Localized and region-based formulations were introduced because global intensity statistics were inadequate under illumination variation and heterogeneous OD appearance [66,67]. Thus, even classical segmentation often combined localization, image normalization, prior geometry, and boundary refinement, rather than relying on a single isolated operator. This same functional decomposition remains evident in learning-based methods, although its components are increasingly learned. Fu et al. [75] retained polar geometry within a convolutional framework; Yi et al. [76] explicitly retained CHT localization before learned refinement; and Bazi et al. [77] implemented localization and refinement through two learned segmentation stages.

Boundary definition remains a central challenge in OD/OC segmentation despite advances in learned representations. Classical active contours were affected by vessel crossings, weak margins, PPA, intensity inhomogeneity, initialization, and the choice of image forces [66–70]. The comparative analysis of Mary et al. [69], for example, showed that different contour formulations addressed different failure modes and that PPA and overexposure remained problematic. Similar concerns appear in recent learning-based work but are handled differently. He et al. [78] added explicit boundary supervision, while Chen et al. [79] reconstructed OD/OC contours from Fourier parameters to preserve shape and structural continuity. Chen et al. introduced the contour-reconstruction component because pixel-level predictions alone did not adequately preserve the expected OD/OC geometry. When the OD/OC boundary is weak or ambiguous, segmentation still requires additional spatial, geometric, or anatomical constraints.

Foundation models further shift the balance by moving image representation from task-specific training to large-scale pretraining. SAM learns broadly transferable segmentation behavior [80]; MedSAM adapts that representation to medical images [81]; and OSAM-Fundus introduces fundus-specific prompt generation and refinement for OD/OC segmentation [82]. RETFound takes a different approach, learning retina-specific representations through self-supervised pretraining [83], which FunduSegmenter subsequently adapts for dense OD/OC segmentation [84]. RETFound itself learns retinal context, including optic-nerve-related anatomical information, from unlabeled retinal images rather than from explicitly programmed descriptors. Nevertheless, FunduSegmenter still requires dedicated adapters, multiscale feature integration, and spatial reconstruction to convert this representation into a segmentation output. Foundation models reduce the need for manually engineered features, but OD/OC segmentation still requires task-specific localization, spatial reconstruction, and boundary delineation.

The more persistent limitation across methodological generations is generalization rather than representation alone. Classical approaches can fail when illumination, pathology, vascular interference, or anatomy violate their predefined assumptions. Data-driven approaches replace many of these assumptions with learned distributions, but they consequently depend on how well the training data represent the target domain. Differences among retinal datasets in scanner type, resolution, illumination, and image appearance can produce substantial domain shift. This concern explains the increasing emphasis on multi-dataset evaluation in Bazi et al. [77], domain adaptation in He et al. [78], cross-camera evaluation in ODFormer [22], one-shot out-of-domain segmentation in OSAM-Fundus [82], and external and domain-generalization testing in FunduSegmenter [84]. Foundation-model pretraining improves transferability, but generalization remains dependent on the target structure and preprocessing pipeline. Zhao et al. [84] reported weaker OC segmentation and found that OD-centered cropping affected external-validation performance. The literature indicates a recurring tradeoff in OD segmentation: spatial and anatomical constraints can improve localization and boundary regularity, but overly restrictive assumptions may reduce robustness when anatomy, pathology, or image layout deviates from the expected model [47–53,66–70, 79,84]. Evaluation is further complicated by differences across retinal datasets in acquisition conditions, pathology, image resolution, annotation protocols, and validation strategies. Bazi et al. [77] addressed this issue with an eight-dataset leave-one-out evaluation. Accordingly, AI-based OD/OC segmentation should be assessed not only by within-dataset performance but also by robustness across datasets, imaging conditions, and challenging anatomical cases. Table 3 summarizes how key OD segmentation principles have evolved from classical image-processing and deformable formulations to contemporary AI-based implementations.

**Table 3.** Evolution of key optic-disc segmentation principles from classical methods to contemporary AI.

| Segmentation principle | Classical implementation | Main limitation | Contemporary implementation | Representative AI studies |
|---|---|---|---|---|
| Appearance representation | Intensity, color, histogram, entropy, and texture descriptors [26–38,62] | Sensitive to illumination, pathology, and manually selected features | Hierarchical and pretrained feature representations | Fu et al. (2018) [75]; Zhou et al. (2023) [83]; Zhao et al. (2026) [84] |
| Geometric prior and localization | Circular/elliptical models, CHT, polar transformation [47–53] | Approximate geometry may fail with irregular discs or PPA | Learned localization combined with geometric or coordinate constraints | Fu et al. (2018) [75]; Yi et al. (2023) [76] |
| Multiscale information | Multiscale filtering and scale-dependent operators [53–61] | Requires predefined scales and parameters | Multiscale feature extraction and fusion within neural architectures | Fu et al. (2018) [75]; Yi et al. (2023) [76]; Wang et al. (2024) [22] |
| Boundary information | Edge operators, matched filters, gradient-based contours [53–61,65–70] | Weak boundaries and vessel interference | Boundary supervision and learned local-feature refinement | He et al. (2024) [78]; Wang et al. (2024) [22] |
| Shape and contour regularization | Hough models, active contours, level sets [47–53,65–70] | Initialization and model-assumption sensitivity | Parameterized contour reconstruction and topology constraints | Chen et al. (2025) [79] |
| ROI/initialization | Seeds, CHT localization, initial contours [47–53,63–70] | Requires reliable initialization | Learned cropping and prompt-conditioned target specification | Bazi et al. (2024) [77]; Kirillov et al. (2023) [80]; Wang et al. (2025) [82] |

| Transfer across datasets | Dataset-specific preprocessing and parameters | Limited portability across acquisition conditions | Domain adaptation, cross-dataset evaluation, and pretrained retinal representations | He et al. (2024) [78]; Bazi et al. (2024) [77]; Wang et al. (2025) [82]; Zhao et al. (2026) [84] |
|---|---|---|---|---|

## 7. Conclusion

This review examines the methodological evolution of optic-disc segmentation, from explicit image processing and deformable models to learned representations, Transformers, and foundation models. Although the representation of image information has changed, core requirements such as localization, boundary delineation, multiscale analysis, and anatomical consistency remain. The reviewed literature shows that these classical principles have not disappeared with AI; rather, they have been reformulated through coordinate transformations, learned localization, boundary supervision, contour reconstruction, attention mechanisms, prompting, and pretrained retinal representations. At the same time, no single representation or prior is universally effective. Strong geometric or spatial constraints can improve segmentation but may reduce robustness when pathology or anatomical variability violates their assumptions. Future progress in OD/OC segmentation should therefore emphasize the flexible integration of learned representations with appropriate anatomical and spatial constraints, together with rigorous evaluation across datasets and imaging conditions. Foundation models offer a promising path to transferable retinal representations, but their value for OD/OC segmentation will ultimately depend on robust boundary delineation, task-specific adaptation, and demonstrated generalization beyond the data used for development. Overall, continued progress will require not only increasingly powerful models but also stronger evidence that their performance remains robust under the variability encountered in real-world retinal imaging.